\documentclass[a4paper,11pt]{article}
\usepackage[toc,page]{appendix}
\usepackage{graphicx}
\usepackage{colortbl}
\usepackage{amsmath}
\usepackage{subfigure}
\usepackage{mathtools}

\DeclarePairedDelimiterX\braket[2]{\langle}{\rangle}{#1 \delimsize\vert #2}
\usepackage[utf8]{inputenc}
\usepackage{float}
\usepackage[normalem]{ulem}
\usepackage{epstopdf}
\usepackage{amsfonts,amsmath,amssymb}
\usepackage{hyperref}

\usepackage{epsfig,multicol,bbm}
\usepackage{color}
\usepackage[dvipsnames]{xcolor}

\definecolor{darkblue}{rgb}{0.0, 0.0, 0.55}
\definecolor{grey}{rgb}{0.57, 0.64, 0.69}
\definecolor{lightbrown}{rgb}{0.71, 0.4, 0.11}

\newcommand{\be}{\begin{equation}}
\newcommand{\ee}{\end{equation}}

\def\ndelta{\delta\hspace{-0.50em}\slash\hspace{-0.05em} }

\newcommand\fverb{\setbox\pippobox=\hbox\bgroup\verb}
\newcommand\fverbit{\egroup\item[\fbox{\unhbox\pippobox}]}

\newbox\pippobox
\begin{document}
\title{\bf Conservation and Integrability in TMG}
\author{ M. R. Setare\thanks{Electronic address: rezakord@ipm.ir}\,,\,S. N. Sajadi\thanks{Electronic address: naseh.sajadi@gmail.com}
\\
\small Department of Science, Campus of Bijar, University of Kurdistan, Bijar, Iran\\
%\small Physics Department and Biruni Observatory, College of Sciences, Shiraz University,
%Shiraz 71454, Iran
}
\maketitle
\begin{abstract}
In this work, following the paper by Romain Ruzziconi and  C\'eline Zwikel \cite{Ruzziconi:2020wrb} we extend the questions of conservation, integrability and
renormalization in Bondi gauge and in GR to the theory of Topological Massive Gravity (TMG).
 We construct the phase space and renormalize the divergences arising within the symplectic structure through a holographic renormalization procedure. We show that the charge expressions are generically finite, conserved and can be made integrable by a field$-$dependent redefinition of
the asymptotic symmetry parameters.
\end{abstract}

\maketitle
\section{Introduction}
For the sake of simplicity the three dimensions space time is an arena for studying various aspects of gravity. Maximally symmetric vacuum solutions of Einstein field equations in three dimensions are characterized by their global properties which encoded in the asymptotic charges. By specifying the boundary conditions which form the asymptotic symmetries, the charges are computed. To perform the boundary conditions and apply the further analyses, it is customary to fix the gauge. Here, we used the Bondi gauge to study asymptotically AdS spacetimes \cite{Bondi:1962px}.
The analysis of asymptotically flat spacetimes at null infinity led to the Bondi mass formula, which states that due to the emission of gravitational waves the mass of the system decreases in time \cite{Bondi:1962px}. The non$-$conservation of charges is an important part to
explain the dynamics of the system which is
related to non$-$integrability of the charges \cite{Barnich:2007bf}, \cite{Wald:1999wa}. Non$-$integrability is commonly considered as an unpleasant property because it implies that
the finite charge expressions rely on the particular path that one chooses to integrate on
the solution space, which is a common feature of a dissipative system \cite{Ruzziconi:2020wrb}.
One way to solve the matter of non$-$integrablity  is to keep the full non$-$integrable
expressions and try to make sense of it. A vital technical result going into this
direction is that the Barnich$-$Troessaert bracket that enables one to derive mathematically
consistent charge algebras for non-integrable charges \cite{Barnich:2010eb}.
This bracket has been employed in
many different contexts and also the associated charge algebras have been
shown to be physically very relevant since they contain all the information concerning the
flux-balance laws of the theory \cite{Barnich:2013axa}.
Progress in understanding the relation between non$-$conservation
and non$-$integrability of the charges has been made. In \cite{Adami:2020ugu}, it was conjectured that no flux passing
through the boundary is equivalent to the existence of a specific section of the phase
space for which the charges are integrable. A natural however non$-$trivial
question that we would like to explore in this work is whether this conjecture is also applicable for
asymptotic boundaries?\\
To get some insights into a attainable well$-$behaved quantum gravity, countless number of works have been devoted to study the lower dimensional models. Notably, studying cosmological Einstein's theory in $3$-dimensions has attracted large attentions. Although the $3D$ cosmological Einstein gravity does not possess any local physical degree of freedom (dof), it has appealing global properties. For example, it possesses the Banados-Teitelboim-Zanelli (BTZ) black hole solutions \cite{ban} or $2D$ boundary CFT with the Brown$-$Henneaux central charge ($c=3l/2G$) dual to the bulk $AdS_3$ spacetime \cite{BrownHenneaux}. Here, the important purpose is to construct a $3D$ gravity theory that also possess a local propagating dof as it has a viable boundary CFT. In this perspective, the Topologically Massive Gravity (TMG) is particularly interesting because it is a renormalizable theory and also describes a \emph{local dynamical} dof \cite{jak}, \cite{Skenderis:2009nt}. The so$-$called shortcoming in cosmological TMG, however, is that the signs of energy of boundary graviton and mass of BTZ black hole are in conflict with each others. For the cosmological TMG, the $AdS_3$ solution is dual to a 2D CFT admitting two copies of Virasoro algebra.\\
When we were working on this paper, the paper \cite{Adami:2021sko} on the subject has been published. In the mentioned paper the authors addressed the quetsion of conservation and integrability of charges  in Bondi gauge and  in TMG. Unlike our paper, they have perturbatively considered the field equations and obtained an approximated solution.\\
The paper is organized as follows: in section \ref{sec2}, we apply the
covariant phase space methods on topological massive gravity in the
Bondi gauge. We discuss the solution
space, renormalize the action principle and the symplectic structure,
obtain the corresponding finite surface charges associated with the asymptotic symmetries. We conclude
with some comments in section \ref{sec3}.

\section{Phase space of TMG}\label{sec2}
Deser, Jackiw and Templeton constructed a $2 + 1$-
dimensional renormalizable dynamical gravity theory with the help of the gravitational
Chern$-$Simons term \cite{jak}. The model is called Topologically Massive Gravity (TMG).
The Lagrangian density of TMG is presented as follows
\begin{align}
L_{TMG}=\dfrac{\sqrt{-g}}{16\pi G}\left[\mathcal{R}-2\Lambda+\dfrac{1}{2 \mu}\epsilon^{\lambda \mu \nu}\left(\Gamma^{\rho}_{\lambda \sigma}\partial_{\mu}\Gamma^{\sigma}_{\rho \nu}+\dfrac{2}{3}\Gamma^{\rho}_{\lambda \sigma}\Gamma^{\sigma}_{\mu \tau}\Gamma^{\tau}_{\nu \rho}\right)\right].
\end{align}
Here $\epsilon^{\lambda \mu \nu}$ is a rank$-$3 tensor. This model
describes a topologically massive graviton possessing a single helicity mode
and acquires asymptotically AdS black hole solutions called
Banados$-$Teitelboim-Zanelli (BTZ)
black holes \cite{ban}. Here we briefly review the phase space of TMG \cite{Compere:2020lrt, rig}.
By an infinitesimal variation of the above Lagrangian we obtain
\begin{equation}
\delta L_{TMG}=\dfrac{\delta L_{TMG}}{\delta g_{\mu \nu}}\delta g_{\mu \nu}+\partial_{\mu}\Theta^{\mu}_{TMG}
\end{equation}
Using the following Euler$-$Lagrange derivatives
\begin{align}
\dfrac{\delta L_{TMG}}{\delta g_{\mu \nu}}=-\dfrac{\sqrt{-g}}{16\pi G}\left(G^{\mu \nu}+\Lambda g^{\mu \nu}+\dfrac{1}{\mu}C^{\mu \nu}\right)
\end{align}
one can find the TMG equations and the canonical TMG presymplectic potential
takes following form
\begin{align}\label{presym}
\Theta^{\mu}_{TMG}=\dfrac{\sqrt{-g}}{16\pi G}\left(g^{\sigma \nu}\delta\Gamma^{\mu}_{\nu \sigma}-g^{\sigma \mu}\delta\Gamma^{\alpha}_{\alpha \sigma}+\dfrac{1}{2\mu}\left(\epsilon^{\lambda \mu \nu}\Gamma^{\rho}_{\lambda \sigma}\delta\Gamma^{\sigma}_{\nu \rho}+\epsilon^{\lambda \sigma \nu}R_{\nu \sigma}{}^{\rho \mu}\delta g_{\lambda \rho}\right)\right)
\end{align}
The TMG theory is invariant under diffeomorphisms which act on the
metric with a standard Lie derivative $\delta_{\xi}g_{\mu \nu}=2\nabla_{(\mu}\xi_{\nu)}$. The weakly$-$vanishing Noether current $S^{\mu}_{\xi}$, is obtained as
\begin{equation}
\dfrac{\delta L_{TMG}}{\delta g_{\mu \nu}}\delta_{\xi}g_{\mu \nu}=\dfrac{\sqrt{-g}}{8\pi G}\left(\nabla_{\mu}G^{\mu \nu}\xi_{\nu}+\dfrac{1}{\mathcal{\mu}}\nabla_{\mu}C^{\mu \nu}\xi_{\nu}\right)+\partial_{\mu}S^{\mu}_{\xi}
\end{equation}
The first and second terms in the
right$-$hand side vanish due to the Bianchi identities which correspond to the Noether identities of the theory. The total derivative term gives $S^{\mu}_{\xi}$,
\begin{equation}
S^{\mu}_{\xi}=2\dfrac{\delta L_{TMG}}{\delta g_{\mu \nu}}=-\dfrac{\sqrt{-g}}{8\pi G}\left(G^{\mu \nu}+\Lambda g^{\mu \nu}+\dfrac{1}{\mu}C^{\mu \nu}\right)\xi_{\nu}.
\end{equation}
The Barnich$-$Brandt codimension
2 form is given by
\begin{equation}
k^{\mu \nu}_{BB, \xi}=\dfrac{1}{2}\delta \phi^{i}\dfrac{\delta}{\delta \phi^{i}_{\nu}}S^{\mu}_{\xi}+\left(\dfrac{2}{3}\partial_{\sigma}\delta \phi^{i}-\dfrac{1}{3}\delta \phi^{i}\partial_{\sigma}\right)\dfrac{\delta}{\delta \phi^{i}_{\nu \sigma}}S^{\mu}_{\xi}-(\mu \leftrightarrow \nu)
\end{equation}
 this 2$-$form in our interesting case takes following form
\begin{align}
k_{BB,\xi}^{\mu \nu}=&k^{\mu \nu}_{E}(\xi)+\epsilon^{\mu \nu \beta}{\cal G}^{(l)}_{\alpha \beta}\bar{\xi}^{\alpha}+\epsilon^{\alpha \nu \beta}{\cal G}^{(l)\mu}{}_{ \beta}\bar{\xi}_{\alpha}+\epsilon^{\mu \alpha \beta}{\cal G}^{\nu}{}_{\beta}\bar{\xi}_{\alpha}+k_{E}^{\mu \nu}(\epsilon \bar{\nabla}\bar{\xi})
\end{align}
where
\begin{align}
k^{\mu \nu}_{E}(\xi)=&\bar{\xi}_{\alpha}\bar{\nabla}^{\mu}h^{\nu \alpha}-\bar{\xi}_{\alpha}\bar{\nabla}^{\nu}h^{\mu \alpha}+\bar{\xi}^{\mu}\bar{\nabla}^{\nu}h-\bar{\xi}^{\nu}\bar{\nabla}^{\mu}h+h^{\mu \alpha}\bar{\nabla}^{\nu}\bar{\xi}_{\nu}-h^{\nu\alpha}\bar{\nabla}^{\mu}\bar{\xi}_{\alpha}+\bar{\xi}^{\nu}\bar{\nabla}_{\alpha}h^{\mu \alpha}\nonumber\\
&-\bar{\xi}^{\mu}\bar{\nabla}_{\alpha}h^{\nu\alpha}+h\bar{\nabla}^{\mu}\bar{\xi}^{\nu},
\end{align}
where
\begin{equation}
{\cal G}^{(l)}_{\mu \nu} =
R^{(l)}_{\mu \nu}-\dfrac{1}{2}\bar{g}_{\mu \nu}R^{(l)}-2\Lambda h_{\mu \nu}\, ,
\end{equation}
with
\begin{align}
R^{(l)}_{\mu \nu} &=\dfrac{1}{2}\left[-\bar{\nabla}^{2}h_{\mu \nu}-\bar{\nabla}_{\mu}\bar{\nabla}_{\nu}h +\bar{\nabla}_{\mu}\bar{\nabla}_{\sigma}h^{\sigma}_{\nu}+\bar{\nabla}_{\nu}\bar{\nabla}_{\sigma}h^{\sigma}_{\mu}\right]  \\
R^{(l)} &=-\bar{\nabla}^{2}h+\bar{\nabla}_{\rho}\bar{\nabla}_{\sigma}h^{\rho \sigma}-2\Lambda h
\end{align}
and $\delta g_{\mu \nu}=h_{\mu \nu}$ is a metric perturbation and $h=g^{\mu \nu}h_{\mu \nu}$.

\subsection{Bondi gauge in three dimensions}
In the following, we will work in the Iyer$-$Wald approach that allows us to renormalize the symplectic structure.

%\tcr{The Bondi gauge in three dimensions has been studied in \cite{Barnich:2010eb} to investigate
%asymptotically AdS$_3$. The analysis was then extended
%to asymptotically locally AdS$_3$ and asymptotically locally flat spacetimes to include the
%boundary structure in the solution space \cite{Ciambelli:2020eba}. We review these results here.}
\subsubsection{Solution space}
Asymptotically AdS spacetimes solving field equation are of the form
\begin{equation}\label{bondimetric}
ds^2=\frac{V}{r}e^{2\beta}du^2-2e^{2\beta}du dr+r^{2}e^{2\varphi}\left(d\phi-U du\right)^2,
\end{equation}
with coordinates $(x^{\mu}) = (u, r, \phi)$. In the metric, $V$, $\beta$ and $U$ are functions of $(u, r, \phi)$ and $\varphi$ is a function of $(u, \phi)$. The Bondi gauge is obtained by requiring the
following three gauge$-$fixing conditions
\begin{equation}\label{eqqbondy}
g_{rr}=0,\;\;\;\;\;g_{r\phi}=0,\;\;\;\;\;g_{\phi \phi}=r^2e^{2\varphi}.
\end{equation}
Applying the TMG's field equations determine the metric for $\Lambda l^2=-1,\mu l\neq 1$ as follows:\\
\begin{align}
\beta(u,r,\phi)=&\beta_{0}(u,\phi)\nonumber\\
U(u,r,\phi)=&U_{0}(u,\phi)+\dfrac{U_{1}(u,\phi)}{r}+\dfrac{U_{2}(u,\phi)}{r^2}\nonumber\\
V(u,r,\phi)=&-\dfrac{U_{2}^{2}(u,\phi)}{r}e^{2\varphi(u,\phi)-2\beta_{0}(u,\phi)}-2 U_{1}(u,\phi)U_{2}(u,\phi)e^{2\varphi(u,\phi)-2\beta_{0}(u,\phi)}+\nonumber\\
&V_{0}(u,\phi)r+V_{1}(u,\phi)r^{2}+V_{2}(u,\phi)r^3
\end{align}
where
\begin{eqnarray}
U_{1}(u,\phi)&=&2e^{2\beta_{0}(u,\phi)-2\varphi(u,\phi)}\partial_{\phi}\beta_{0}(u,\phi),\\
U_{2}(u,\phi)&=&-e^{2\beta_{0}(u,\phi)-2\varphi(u,\phi)}N(u,\phi),\\
V_{1}(u,\phi)&=&\dfrac{1}{3U_{2}e^{4\varphi+2\beta_{0}}+2\mu r^{2}e^{4\beta_{0}+3\varphi}}\left[(12U_{2}\partial_{u}\varphi-12U_{2}\partial_{u}\beta_{0}+6\partial_{\phi}(U_{0}U_{2})\right.  \nonumber\\
&&\left.
+12U_{2}U_{0}(\partial_{\phi}\varphi-\partial_{\phi}\beta_{0})+6\partial_{u}U_{2})e^{4\varphi+2\beta_{0}}-4\mu r^2(\partial_{u}\varphi+U_{0}\partial_{\phi}\varphi+\partial_{\phi}U_{0})e^{4\beta_{0}+3\varphi}\right. \nonumber \\
&&\left.+(6V_{0}\partial_{\phi}\beta_{0}+3\partial_{\phi}V_{0})e^{4\beta_{0}+2\varphi}+48\partial_{\phi}\beta_{0}e^{6\beta_{0}}(2(\partial_{\phi}\beta_{0})^{2}-\partial_{\phi}\varphi \partial_{\phi}\beta_{0}+\partial^{2}_{\phi^{2}}\beta_{0})\right]\\
V_{2}(u,\phi)&=&\dfrac{1}{2\mu l^{2}r^{3} e^{4\beta_{0}+3\varphi}}\left[-2l^{2}e^{4\varphi+2\beta_{0}}[\partial_{\phi}(U_{2}U_{0})+2U_{2}U_{0}\partial_{\phi}\varphi+2U_{2}\partial_{u}\varphi-\right. \nonumber \\
&&\left. 2U_{2}U_{0}\partial_{\phi}\beta_{0}-2U_{2}\partial_{u}\beta_{0}+\partial_{u}U_{2}-\dfrac{1}{2}U_{2}V_{1}]-l^2e^{4\beta_{0}+2\varphi}[2V_{0}\partial_{\phi}\beta_{0}+\partial_{\phi}V_{0}]\right.\nonumber\\
&&\left. -2\mu r^{3}e^{6\beta_{0}+3\varphi}+8l^2\partial_{\phi}\beta_{0}e^{6\beta_{0}}[-2(\partial_{\phi}\beta_{0})^{2}+\partial_{\phi}\varphi \partial_{\phi}\beta_{0}-\partial^{2}_{\phi^2}\beta_{0}]\right]
\end{eqnarray}
We also have two constrained field equations which determine the time evolution of $N$ and $V_{0}$ (see appendix \ref{appcons}).
The expansion of the metric around the AdS spacetime is
\begin{eqnarray}
\dfrac{V}{r}&=&\left[-\dfrac{r^2 e^{2\beta_{0}}}{l^2}-2r\left(U_{0}\partial_{\phi}\varphi +\partial_{u}\varphi +\partial_{\phi}U_{0}\right)+V_{0}+\dfrac{4N\partial_{\phi}\beta_{0}e^{2\beta_{0}-2\varphi}}{r}-\dfrac{N^2e^{2\beta_{0}-2\varphi}}{r^2}\right]\nonumber\\
&&+\dfrac{2e^{-\varphi}}{\mu r}\left[e^{2\beta_{0}-2\varphi}\left(4\partial_{\phi}\beta_{0}\partial^{2}_{\phi}\beta_{0}+8(\partial_{\phi}\beta_{0})^3-
4\partial_{\phi}\varphi (\partial_{\phi}\beta_{0})^2\right)+V_{0}\partial_{\phi}\beta_{0}-NU_{0}\partial_{\phi}\varphi -\right.\nonumber\\
&&\left. U_{0}\partial_{\phi}N-N\partial_{u}\varphi -2N\partial_{\phi}U_{0}-\partial_{u}N+\dfrac{1}{2}\partial_{\phi}V_{0}\right]+\mathcal{O}(\mu^{-2}).
\end{eqnarray}
The expression inside the first bracket is the solution of Einstein gravity \cite{Ciambelli:2020ftk}, and the second bracket comes from the TMG. If the $ \partial_{u}N $ equals to the equation (23)\footnote{$\partial_{u}N+N\partial_{u}\varphi =e^{2\beta_{0}-2\varphi}\left(4\partial_{\phi}\beta_{0}\partial^{2}_{\phi}\beta_{0}+8(\partial_{\phi}\beta_{0})^3-
4\partial_{\phi}\varphi (\partial_{\phi}\beta_{0})^2\right)+V_{0}\partial_{\phi}\beta_{0}-NU_{0}\partial_{\phi}\varphi -U_{0}\partial_{\phi}N-2N\partial_{\phi}U_{0}+\dfrac{1}{2}\partial_{\phi}V_{0}$} of \cite{Ciambelli:2020ftk}, then the second bracket equals zero. Also, the Cotton tensor for our solution does not equal zero.
The solution space is paramatrized
by five functions which three of them $(\beta_0, U_0, \varphi)$ characterize the induced boundary metric through
\begin{equation}\label{boundraymetric}
\bar{\gamma}_{a b}dx^{a}dx^{b}=\lim_{r\rightarrow \infty}\left(\dfrac{1}{r^2}ds^2\right)=\left(-\dfrac{e^{4\beta_{0}(u,\phi)}}{l^2}+U_{0}^2e^{2\varphi(u,\phi)}\right)du^2-2U_{0}e^{2\varphi}du d\phi+e^{2\varphi}d\phi^2\;,
\end{equation}
where $(x^a) = (u, \phi)$ are the coordinates on boundary. The other two functions $V_{0}$ and $N$ encode
the bulk information on the mass and the angular momentum.

\subsection{Residual symmetries}
The residual gauge diffeomorphisms $\xi=\xi^{u}\partial_{u}+\xi^{\phi}\partial_{\phi}+\xi^{r}\partial_{r}$  preserving the Bondi gauge fixing (\ref{eqqbondy}) have to satisfy the conditions
\begin{equation}
\mathcal{L}_{\xi}g_{rr}=\mathcal{L}_{\xi}g_{r\phi}=0,\;\;\;\;g^{\phi \phi}\mathcal{L}_{\xi}g_{\phi \phi}=2w(u,\phi),
\end{equation}
which the explicit solutions of them are given by
\begin{align}\label{killingvec}
\xi^{u}=&f\nonumber\\
\xi^{\phi}=&Y-\dfrac{1}{r}\partial_{\phi}f e^{2\beta_{0}-2\varphi}\nonumber\\
\xi^{r}=&r(\partial_{\phi}f U_{0}+w-\partial_{\phi}Y-f\partial_{u}\varphi -Y\partial_{\phi}\varphi)+\partial_{\phi} f U_{1}+\dfrac{\partial_{\phi}f U_{2}}{r}+\nonumber\\
&e^{2\beta_{0}-2\varphi}\left(\partial^{2}_{\phi}f+2\partial_{\phi}\beta_{0}\partial_{\phi}f-\partial_{\phi}f \partial_{\phi}\varphi\right).
\end{align}
So, the residual diffeomorphisms are encoded in $f, Y, w$ which arbitrary functions of the boundary coordinates. Using the modified Lie bracket
\begin{equation}\label{eqcomut}
[\xi_{1},\xi_{2}]_{\ast}=[\xi_{1},\xi_{2}]-\delta_{\xi_{1}}\xi_{2}+\delta _{\xi_{2}}\xi_{1},
\end{equation}
 these vector fields satisfy the commutation relations $[\xi(f_1,Y_1,w_1),\xi(f_2,Y_{2},w_2)]_{\ast}=\xi(f_{12},Y_{12},w_{12})$ where
\begin{align}\label{eqqcomut}
f_{12}=&f_{1}\partial_{u}f_{2}+Y_{1}\partial_{\phi}f_{2}-\delta_{\xi_{1}}f_{2}-(1\leftrightarrow 2)\nonumber\\
Y_{12}=&f_{1}\partial_{u}Y_{2}+Y_{1}\partial_{\phi}Y_{2}-\delta_{\xi_{1}}Y_{2}-(1\leftrightarrow 2)\nonumber\\
w_{12}=&-\delta_{\xi_{1}}w_{2}-(1\leftrightarrow 2).
\end{align}

Under the infinitesimal residual gauge diffeomorphisms ($\mathcal{L}_{\xi}g_{a b}=\delta_{\xi}g_{a b}$), the boundary structure
transforms as
\begin{align}\label{eqtrans}
\mathcal{L}_{\xi}g_{\phi \phi}=\delta_{\xi}g_{\phi \phi},\;\;\;\rightarrow\;\;\; \delta_{\xi}\varphi =&w\nonumber\\
\mathcal{L}_{\xi}g_{u r}=\delta_{\xi}g_{u r},\;\;\;\rightarrow\;\;\; \delta_{\xi}\beta_{0}=&(f\partial_{u}+Y\partial_{\phi})\beta_{0}+\left(\dfrac{1}{2}\partial_{u}-\dfrac{1}{2}\partial_{u}\varphi+U_{0}\partial_{\phi}\right)f-\dfrac{1}{2}\left(\partial_{\phi}Y+Y\partial_{\phi}\varphi-w\right)\nonumber\\
\mathcal{L}_{\xi}g_{\phi u}=\delta_{\xi}g_{\phi u},\;\;\;\rightarrow\;\;\;\delta_{\xi}U_{0}=&f\partial_{u}U_{0}+Y\partial_{\phi}U_{0}-U_{0}\partial_{\phi}Y-\partial_{u}Y+\dfrac{1}{l^2}e^{4\beta_{0}-2\varphi}\partial_{\phi}f+U_{0}(\partial_{u}f+U_{0}\partial_{\phi}f)
\end{align}
and the variations of the $N$ and $V_{0}$ can be found in \cite{Ciambelli:2020eba}.

In Bondi gauge, the conditions
\begin{equation}\label{eqdir}
\beta_{0}=0,\;\;\;\;\; U_{0}=0,\;\;\;\;\; \varphi=0,
\end{equation}
corresponds to the Brown$-$Henneaux boundary condition, i.e. the induced boundary metric (\ref{boundraymetric}) is flat. These lead to the additional conditions as follows
\begin{equation}\label{eqdir2}
\partial_{u}f=\partial_{\phi}Y,\;\;\;\;\partial_{u}Y=\dfrac{1}{l^2}\partial_{\phi}f\;\;\;\;\;\;w=0.
\end{equation}

\subsection{Renormalization of the phase space}
By evaluating the radial component of the TMG presymplectic potential (\ref{presym}) on the solution space, we obtain some $\mathcal{O}(r^{2})$, $\mathcal{O}(r)$ terms that diverge when $r\rightarrow \infty$. Also, there is $\mathcal{O}(\ln(r))$ divergence which comes from the Chern$-$Simons part of on$-$shell action. Furthermore, the presymplectic potential admits some $\mathcal{O}(l^{2})$ terms that make an prevention to take the flat limit $l\rightarrow \infty$. Therefore, to remove these divergences, we
should add some counter-terms which we show below the exact form of them. In asymptoticaly $AdS_3$ spacetimes, the minimal action principle in Bondi gauge that satisfies the mentioned requirements is given by \cite{deHaro:2000vlm},\cite{Ruzziconi:2020wrb},\cite{Compere:2020lrt}
\begin{align}\label{totalact}
S=&\dfrac{1}{16\pi G}\int_{M}\sqrt{-g}\left[\mathcal{R}-2\Lambda+\dfrac{1}{2 \mu}\epsilon^{\lambda \mu \nu}\left(\Gamma^{\rho}_{\lambda \sigma}\partial_{\mu}\Gamma^{\sigma}_{\rho \nu}+\dfrac{2}{3}\Gamma^{\rho}_{\lambda \sigma}\Gamma^{\sigma}_{\mu \tau}\Gamma^{\tau}_{\nu \rho}\right)\right]d^{3}x+\int_{\partial M}a_{1}L_{GHY}d^{2}x\nonumber\\
&+\int_{\partial M}a_{2}L_{ct}d^{2}x+\int_{\partial M}a_{3}L_{0}d^{2}x+\int_{\partial M}a_{4}L_{b}d^{2}x+\int_{\partial M}a_{5}L_{\mathcal{R}}d^{2}x+\dfrac{1}{\mu}\int_{\partial M}a_{6}L_{ccs}d^{2}x\nonumber\\
&+\dfrac{1}{\mu}\int_{\partial M}a_{7}L_{lccs}\log(r)d^{2}x
\end{align}
The first term is the TMG bulk action and the second term is the Gibbons$-$Hawking boundary
term as
\begin{equation}
L_{GHY}=\dfrac{1}{8\pi G}\sqrt{-\gamma}\gamma^{\mu \nu}n_{\nu ;\mu}
\end{equation}
where $n_{\mu}=\dfrac{1}{\sqrt{g^{rr}}}\delta^{r}_{\mu}=\sqrt{\dfrac{re^{2\beta_{0}(u,\phi)}}{-V(u,r,\phi)}}\delta^{r}_{\mu}$ and deteminent of induced metric $\gamma=\left. rVe^{2(\varphi+\beta)}\right\vert_{r=R}$. The induced metric is $\gamma_{a b}$ and obtained by taking $r=R=constant$ as
\begin{equation}
ds_{B}^2=\dfrac{V(u,\phi)}{R}e^{2\beta_{0}}du^2+R^{2}e^{2\varphi}\left(d\phi-U(u,\phi)du\right)^{2}.
\end{equation}
The third term is the counter$-$term as
\begin{equation}
L_{ct}=-\dfrac{1}{8\pi G l}\sqrt{-\gamma}=-\dfrac{1}{8\pi G l}e^{\varphi+\beta}\sqrt{-R V}.
\end{equation}
The fourth term is a corner Lagrangian
\begin{equation}\label{eqL0}
L_{0}=-\dfrac{1}{8\pi G}\sqrt{-\gamma}D_{a}v^{a},\;\;\;\;\; v^{a}=\dfrac{R e^{\varphi}}{\sqrt{-\gamma}}(\delta^{a}_{u}+U\delta^{a}_{\phi}).
\end{equation}
The vector field $v^{a}$ appeared in (\ref{eqL0}) is tangent to the leaves of the foliation and satisfies the two properties\cite{Detournay:2014fva}, \cite{Compere:2020lrt}
\begin{equation}
v^{a}\gamma_{a b}v^{b}=-1,\;\;\;\;\;\lim_{r\rightarrow \infty}\left(\dfrac{1}{r}\gamma_{a b}v^{a}\right)=-\dfrac{e^{2\beta_{0}}}{l}\delta^{u}_{a}.
\end{equation}
The next term in (\ref{totalact}) is a kinetic term for the vector ($v^{a}$),
\begin{equation}
L_{b}=\dfrac{l}{16\pi G}\sqrt{-\gamma}\left(D_{a}v^{a}\right)^{2}.
\end{equation}
The next term in (\ref{totalact}) is the Gauss$-$Bonnet term
\begin{equation}
L_{\cal{R}}=-\dfrac{l}{16\pi G}\sqrt{-\gamma}\mathcal{R}[\gamma].
\end{equation}
%\tcr{
%\begin{equation}
%L_{bcs}=\sqrt{-\gamma}\epsilon^{\lambda \mu \nu}\left(n_{\mu}\Gamma^{\rho}_{\lambda \sigma}\delta \Gamma^{\sigma}_{\nu \rho}+n_{\sigma}R_{\nu \mu}{}^{\rho \sigma}\delta g_{\lambda \rho}\right)
%\end{equation}}

Finally, the last terms in (\ref{totalact}) is a term to remove the divergence correspond to $\mathcal{O}(R)$ and $\log(R)$ that are coming from the Chern-Simons term of TMG
\begin{equation}
L_{ccs}=\sqrt{-\bar{\gamma}}\bar{\epsilon}^{\lambda \sigma}N_{\rho}\bar{\Gamma}^{\rho}_{\lambda \sigma},\;\;\;\;\; N_{\rho}=-\dfrac{l r}{2}e^{-\Phi(u,\phi)}\delta^{u}_{ \rho}
\end{equation}
\begin{equation}
L_{lccs}=\partial_{a}\bar{m}^{a},\;\;\;\;\; \bar{m}^{a}=\partial_{\phi}\beta_{0}(\delta^{a}_{u}+U_{0}\delta^{a}_{\phi})
\end{equation}

The explicit terms of the Lagrangians are given in appendix \ref{appexpl}.
Now we provide some details on how to obtain the values of the coefficients (\ref{totalact})
of the different terms in the action.
\begin{itemize}
\item By evaluating the action (\ref{totalact}) on$-$shell, divergences in $\mathcal{O}(R^{2})$ arise. The $R^2$-
divergences by imposing $2a_1-a_2 = 1$ are removed, while the $R$-divergences are
removed by $ 2a_1 -a_2-a_3 = 0$ and $a_{6}=1$.

\item The $\ln(R)$-divergences are removed by imposing: $a_{7}=1$.

\item The action (\ref{totalact}) also exhibits some terms in $\mathcal{O}(l^2)$ which can be
eliminated by imposing $a_5 =-a_1$ and $2a_1-a_2-a_4 = 0$.

\item By using the Dirichlet boundary conditions (\ref{eqdir}) and $a_1=1$, the action is stationary on solutions space.
\end{itemize}

Using all the constraints, we obtain $a_{i}=1$. Sending the cut$-$off to infinity,
$R\rightarrow \infty$, the expression of the renormalized action (\ref{totalact}) is given by
\begin{eqnarray}
S_{ren}&=&\dfrac{1}{16\pi G}\int d^{2}x\left(-V_{0}e^{\varphi}+4e^{2\beta_{0}-\varphi}(-\partial^{2}_{\phi^{2}}\beta_{0}+\partial_{\phi}\beta_{0}\partial_{\phi}\varphi-2(\partial_{\phi}\beta_0)^2)\right)
%\nonumber\\
%&&+\dfrac{1}{\mu}\left(V_{0}U_{0}e^{2\varphi -2\beta_{0}}+U_{0}\partial^{2}_{\phi^{2}}\beta_{0}-U_{0}\partial^{2}_{\phi^{2}}\varphi +\partial^{2}_{u \phi}\beta_{0}-\partial^{2}_{ \phi}U_{0}-\partial^{2}_{u\phi}\varphi -\partial_{u}\varphi \partial_{\phi}\beta_{0}-\partial_{\phi}\varphi U_{0}\partial_{\phi}\beta_{0}\right.\nonumber\\
%&&\left. +2U_{0}(\partial_{\phi}\beta_{0})^{2}+\partial_{\phi}U_{0}\partial_{\phi}\beta_{0}-2U_{0}\varphi \partial^{2}_{\phi^2}\beta_{0}+3U_{0}\partial^{2}_{\phi^2}\beta_{0}+2\beta_{0}U_{0}\partial^{2}_{\phi^2}\beta_{0}+
%2U_{0}\varphi \partial^{2}_{\phi^2}\varphi -2\beta_{0}U_{0}\partial^{2}_{\phi^2}\varphi \right.\nonumber\\
%&&\left. -U_{0}\partial^{2}_{\phi^2}\varphi -U_{0}\partial_{\phi}\beta_{0}\partial_{\phi}\varphi -\partial_{\phi}U_{0}\partial_{\phi}\varphi-\partial_{u}\varphi\partial_{\phi}\varphi +2(\beta_{0}-\varphi)\partial^{2}_{u\phi}\beta_{0}+2(\beta_{0}-\varphi)(\partial^{2}_{u\phi}\beta_{0}-\partial^{2}_{\phi^2}U_{0}-\right.  \nonumber\\
%&&\left. \partial^{2}_{u\phi}\varphi -\partial_{\phi}\varphi \partial_{\phi}U_{0}+4\partial_{\phi}U_{0}\partial_{\phi}\beta_{0})
%+(\partial_{\phi}\beta_{0}-\partial_{\phi}\varphi)(\partial_{\phi}U_{0}-\partial_{u}\varphi)+2\partial_{\phi}\beta_{0}\partial_{u}\beta_{0}+6U_{0}\partial_{\phi}\beta_{0}^2
%\right)\nonumber\\
-\Gamma_{bulk}(R_{0})\;,
\end{eqnarray}
where $\Gamma_{bulk}$ is the finite contribution of the bulk action evaluated on its lower
bound.
\subsection{Renormalization of the symplectic structure}
By inserting the Bondi metric (\ref{bondimetric}) into (\ref{presym}), the
radial component of the presymplectic potential gives some divergences.
The counter$-$terms to remove the $\mathcal{O}(r^2)$ and $\mathcal{O}(l^2)$ divergences are similar to those used in (\ref{totalact}) to renormalize the action.
We define the renormalized presymplectic potential as \cite{Ruzziconi:2020wrb},\cite{Compere:2020lrt}
\begin{align}\label{eqpresem}
\Theta^{r}_{ren}=&\Theta^{r}_{TMG}+\delta L_{GHY}+\delta L_{ct}+\delta L_{0}+\delta L_{b}+\delta L_{\mathcal{R}}+\dfrac{1}{\mu}\delta L_{ccs}\nonumber\\
&-\partial_{a}\Theta^{a}_{0}-\dfrac{1}{2}r\partial_{a}\bar{\Theta}^{a}_{0}
\end{align}
where $g$ is a solution, $\delta g$ a perturbation around it.
The first line is the presymplectic potential prescribed by the renormalized action (\ref{totalact}).
%The second line fixes the $d$-exact ambiguity appearing in (1.3) as
%\begin{equation}
%Y^{a r}=\Theta^{a}_{0}+\dfrac{1}{2}r\bar{\Theta}^{a}_{0}.
%\end{equation}
%Here, $\Theta^{a}_{0}$ is the presymplectic potential associated with the corner Lagrangian (3.24),
%i.e $\delta L_{0}=\partial_{a}\Theta^{a}_{0}$.
 To understand the second term in (\ref{eqpresem}), we define the following quantities induced on the boundary
\begin{equation}
\bar{v}^{a}=\lim_{r\rightarrow \infty}(r v^{a})=le^{-2\beta_{0}}(\delta^{a}_{u}+U_{0}\delta^{a}_{\phi}),\;\;\;\;\; \bar{\gamma}_{a b}=\lim_{r\rightarrow \infty}\left(\dfrac{1}{r^2}\gamma_{a b}\right)
\end{equation}
and $\bar{\gamma}=det(\bar{\gamma}_{a b})$ and $\bar{D}_{a}$ the determinant and the covariant derivative associated with the induced boundary metric $\bar{\gamma}_{a b}$, respectively.
The term $\bar{\Theta}^{a}_{0}$ is the presymplectic potential of the boundary Lagrangian
\begin{equation}\label{eqboundray}
\bar{L}_{0}=-\dfrac{1}{8\pi G}\sqrt{-\bar{\gamma}}\bar{D}_{a}\bar{v}^{a},\;\;\;\;\; \delta \bar{L}_{0}=\partial_{a}\bar{\Theta}^{a}_{0}=-\dfrac{1}{8\pi G}\sqrt{-\bar{\gamma}}\bar{D}_{a}\delta \bar{v}^{a}
\end{equation}
where the variation is taken with respect to $\bar{v}^{a}$ by keeping the boundary metric $\bar{\gamma}_{a b}$ fixed (i.e. $\bar{\gamma}_{a b}$ is seen as a background). By using above setup one can obtain the presymplectic potential and the associated presymplectic current which explicitly provided as equation (\ref{thetaren}) and (\ref{eqwren}) in Appendix (\ref{appexpl1}). When we impose the
conditions (\ref{eqdir}), at leading order this expression vanishes. Hence, the associated charges are conserved and the variational principle (\ref{totalact}) is stationary on solutions.

%\tcb{
%This divergence should be removed by adding the suitable counter term..
%\begin{align}
%\dfrac{R}{\mu}\left(\dfrac{1}{2}U_{0}\partial_{\phi}\delta U_{0}+\dfrac{1}{2}\partial_{u}\delta U_{0}+\delta U_{0}(\partial_{\phi}\varphi U_{0}-\partial_{\phi}\beta_{0}U_{0}+\partial_{u}\varphi +\dfrac{1}{2}\partial_{\phi}U_{0}-\partial_{u}\beta_{0})\right)e^{2\varphi -2\beta_{0}}
%\end{align}}

\subsection{Integrability and charge algebra}
In this section, we talk about the renormalized charges and present a particular slicing of the
phase space in which they are integrable. The renormalized co$-$dimension 2 form can be derived using
\begin{equation}
\partial_{a}k^{r a}{}_{ren,\xi}[g;\delta g]=w^{r}{}_{ren}[g;\delta_{\xi}g,\delta g]
\end{equation}
where $w^{r}{}_{ren}[g;\delta_{1}g,\delta_{2}g]$ given in (\ref{eqwren}). We obtain the infinitesimal charges by integration on $S^{1}_{\infty}$,
\begin{equation}\label{eqintg}
\ndelta Q_{\xi}=\int_{0}^{2\pi}d\phi \;k^{u r}{}_{ren,\xi},
\end{equation}
where $\ndelta$ indicates that the charge is not integrable. The explicite expression has been provided as equation (\ref{eqchar}) in Appendix (\ref{appexpl1}).

By using the renormalization procedure (\ref{eqpresem}), the charges are finite.
The charges (\ref{eqchar}) seem to be non$-$integrable. This obstruction for integrability can be cured by performing field$-$dependent redefinitions of the symmetry parameters. In our case, we do the redefinition
\begin{equation}\label{eqred}
\tilde{f}=fe^{2\beta_{0}-\varphi},\;\;\;\;\tilde{Y}=Y-U_{0}f,\;\;\;\; \tilde{w}=w
\end{equation}
where $\tilde{f}$, $\tilde{Y}$ and $\tilde{w}$ are taken to be field$-$independent, i.e. $\delta \tilde{f}=\delta\tilde{Y}=\delta\tilde{w}=0$. In terms of these parameters, the commutation relations (\ref{eqqcomut}) become
$[\xi(\bar{f}_{1},\bar{Y}_{1},
\bar{w}_{1}),\xi(\bar{f}_{2},\bar{Y}_{2},
\bar{w}_{2})]_{\ast}=\xi(\bar{f}_{12},\bar{Y}_{12},
\bar{w}_{12})$ with
\begin{align}\label{eqcond47}
\bar{w}_{12}&=0\nonumber\\
\bar{f}_{12}&=\bar{Y}_{1}\partial_{\phi}\bar{f}_{2}+\bar{f}_{1}\partial_{\phi}\bar{Y}_{2}-(1\leftrightarrow 2)\nonumber\\
\bar{Y}_{12}&=\bar{Y}_{1}\partial_{\phi}\bar{Y}_{2}+\dfrac{1}{l^2}\bar{f}_{1}\partial_{\phi}\bar{f}_{2}-(1\leftrightarrow 2).
\end{align}
The redefinition (\ref{eqred}) gives the charges (\ref{eqchar}) integrable. We have explicitly $\ndelta Q_{\xi}\equiv \delta Q_{\xi}$
with
\begin{align}\label{eqinch}
\delta Q_{\xi}&=\dfrac{1}{8\pi G}\int_{0}^{2\pi}d\phi\left[\tilde{Y}\delta(Ne^{\varphi})+\tilde{f}\delta\left(\dfrac{V_{0}}{2}e^{2\varphi-2\beta_{0}}+4(\partial_{\phi}\beta_{0})^{2}-2\partial_{\phi}\beta_{0}\partial_{\phi}\varphi+\dfrac{1}{2}(\partial_{\phi}\varphi)^{2}+\partial^{2}_{\phi}(2\beta_{0}-\varphi)\right)\right]\nonumber\\
&+\dfrac{1}{4\pi G\mu}[\tilde{Y}\delta(V_{0}e^{-2\beta_{0}+2\varphi}+7(\partial_{\phi}\beta_{0})^{2}-3\partial_{\phi}\varphi \partial_{\phi}\beta_{0}+(\varphi-2\beta_{0})\partial_{\phi}\beta_{0}\partial_{\phi}\varphi-2(\varphi-2\beta_{0})(\partial_{\phi}\beta_{0})^2+\nonumber\\
&(\varphi-\beta_{0})\partial_{\phi}^{2}(2\beta_{0}-\varphi)+2\partial^{2}_{\phi}\beta_{0})+\tilde{w}\delta(\partial_{\phi}\beta_{0}+\partial_{\phi}\beta_{0}(\varphi-2\beta_{0}))+\tilde{f}\delta(-\partial_{\phi}U_{0}\partial^{2}_{\phi}e^{-2\beta_{0}+\varphi}+\partial^{2}_{\phi}U_{0}(\varphi-\beta_{0})\nonumber\\
&-\varphi\partial_{\phi}U_{0}\partial_{\phi}\beta_{0})]
\end{align}
Integrating the expression (\ref{eqinch}) and using conditions (\ref{eqdir2}) gives the finite charge as
\begin{equation}\label{eqchin}
Q_{\xi}[g]=\dfrac{1}{16\pi G}\int_{0}^{2\pi}d\phi \left[\tilde{Y}\tilde{N}+\tilde{f}\tilde{V_{0}}\right]
\end{equation}
where
\begin{align}
\tilde{N}=&Ne^{\varphi}+\frac{2}{\mu}(V_{0}e^{-2\beta_{0}+2\varphi}+7(\partial_{\phi}\beta_{0})^{2}-3\partial_{\phi}\varphi \partial_{\phi}\beta_{0}+(\varphi-2\beta_{0})\partial_{\phi}\beta_{0}\partial_{\phi}\varphi-2(\varphi-2\beta_{0})(\partial_{\phi}\beta_{0})^2+\nonumber\\
&(\varphi-\beta_{0})\partial_{\phi}^{2}(2\beta_{0}-\varphi)+2\partial^{2}_{\phi}\beta_{0}),\;\;\;\; \tilde{V}_{0}=\dfrac{V_{0}}{2}e^{2\varphi-2\beta_{0}}+4(\partial_{\phi}\beta_{0})^{2}-
2\partial_{\phi}\beta_{0}\partial_{\phi}\varphi+\dfrac{1}{2}(\partial_{\phi}\varphi)^{2}+\nonumber\\&\partial^{2}_{\phi}(2\beta_{0}-\varphi)+\frac{2}{\mu}(-\partial_{\phi}U_{0}\partial^{2}_{\phi}e^{-2\beta_{0}+\varphi}+\partial^{2}_{\phi}U_{0}(\varphi-\beta_{0})
-\varphi\partial_{\phi}U_{0}\partial_{\phi}\beta_{0}).
\end{align}
As can be seen there are only two independent charges.\\
According to the theorem of the covariant phase space method, the algebra of charges is the same as the algebra of symmetry generators up to central terms \cite{Iyer:1994ys}. By using the explicit form of the charges and also using Dirichlet boundary conditions (\ref{eqdir}) and (\ref{eqdir2}), one can compute the charge algebra as follows \cite{leewald}, \cite{Barnich:2011mi}, \cite{Dengiz:2020fpe}
\begin{equation}
\delta_{\xi_{2}}Q_{\xi_{1}}\equiv \lbrace Q_{\xi_{1}},Q_{\xi_{2}}\rbrace =Q_{\left[ \xi_{1},\xi_{2}\right]_{\ast} }+C_{\xi_{1},\xi_{2}},
\end{equation}
where $\left[\xi_{1},\xi_{2}\right]_{\ast} $ is given by (\ref{eqcond47}) and $ C_{\xi_{1},\xi_{2}} $ is the central extension
\begin{equation}\label{eqcent}
C_{\xi_{1},\xi_{2}}=\dfrac{1}{8\pi G}\int_{0}^{2\pi}d\phi \left(\partial^{2}_{\phi}\tilde{f}_{1}\partial_{\phi}\tilde{Y}_{2}-
\partial^{2}_{\phi}\tilde{f}_{2}\partial_{\phi}\tilde{Y}_{1}\right)+
\dfrac{1}{16\pi G \mu l^2}\int_{0}^{2\pi}d\phi \left(\partial_{\phi}\tilde{f}_{2}\partial^{2}_{\phi}\tilde{f}_{1}-
\partial_{\phi}\tilde{f}_{1}\partial^{2}_{\phi}\tilde{f}_{2}\right).
\end{equation}
We should mention that this is the first time that the central extension with the gravitational anomaly of TMG appears for such weak boundary conditions we have considered in this paper.\\
As can be seen from (\ref{eqcent}), there are two kinds of central extension, the first term arises from the Einstien term and the other one from the CS term and appears in the Virasoro part. Defining
\begin{equation}
\tilde{f}=\dfrac{l}{2}\left(Y^{+}+Y^{-}\right),\hspace{1cm}\tilde{Y}=\dfrac{1}{2}\left(Y^{+}-Y^{-}\right)
\end{equation}
and using the decomposition in modes $Y^{\pm}=\Sigma Y_{m}^{\pm}l^{\pm}_{m}$, $l^{\pm}_{m}=e^{\pm im x^{\pm}}$ and writing $L_{\pm}=Q_{\xi(l^{\pm}_{m})}$, the charge algebra becomes
\begin{equation}
i\lbrace L^{\pm}_{m},L^{\pm}_{n}\rbrace =(m-n)L^{\pm}_{m+n}-\dfrac{c^{\pm}}{12}m^{3}\delta_{m+n,0},\hspace{0.5cm}\lbrace L_{m}^{\pm},L_{n}^{\mp}\rbrace =0,
\end{equation}
where $c^{\pm}=\dfrac{3l}{2G}\pm\dfrac{3}{2\mu G}$ which is the Brown$-$Henneaux central charges.

% However, one expects from the
%arguments presented in that the maximal number of independent charges in this
%context is three. As already suggested in, this confirms that three-dimensional
%Bondi gauge is not the most general framework to study the maximal phase space of the
%theory.

\section{Conclusion}\label{sec3}

In this work, we construct the maximal asymptotic symmetry algebra that one will get in
topological massive 3D gravity by imposing partial gauge fixing on the components of the metric
and very mild falloffs. After a renormalization procedure of the action and the
symplectic structure involving covariant counter$-$terms, we obtain finite charge
expressions that we present integrable through a field$-$dependent redefinition of the symmetry
parameters. Then, we find the appropriate field$-$dependent redefinition of the parameters to exhibit the charges integrable. Finally, we have obtained the charge algebra which involves the Brown$-$Henneaux central extension in asymptotically AdS$_3$ spacetimes. This is the first time that the central extension with the gravitational anomaly of TMG appears for such weak boundary conditions we have considered in this paper.
\section{Acknowledgments}
We would like to thank Romain Ruzziconi and Kostas Skenderis for discussions and comments.
\appendix
\section{Constrained field equations}\label{appcons}
The constrained components of field equations are presented as follows:
\begin{align}
E_{u\phi}=&\dfrac{1}{8\mu r^3}[6r^5\partial_{r}U(u,r,\phi)e^{-6\beta_{0}+3\varphi}(2\partial_{r}U(u,r,\phi)+r\partial^{2}_{r}U(u,r,\phi))-4\mu r^{4}e^{-4\beta_{0}+2\varphi}(3\partial_{r}U(u,r,\phi)\nonumber\\
&+r\partial^{2}_{r}U)
(6r\partial_{r}V-3r^2\partial^{2}_{r}V-6V+r^3\partial^{3}_{r}V)e^{-4\beta_{0}+\varphi}-8\mu r^{2}\partial_{\phi}\beta_{0}e^{-2\beta_{0}}]=0,
\end{align}
and
\begin{align}
E_{r\phi}&=\dfrac{1}{8\mu r^3}[-6r^6 \partial_{r}U e^{-6\beta_{0}+3\varphi}(U\partial^{2}_{r\phi}U+\partial^{2}_{u r}U+\partial_{r}U(-U\partial_{\phi}\beta_{0}+\partial_{u}\varphi+U\partial_{\phi}\varphi-2\partial_{u}\beta_{0}+2\partial_{\phi}U))\nonumber\\
&4\mu r^{5}e^{-4\beta_{0}+2\varphi}(U\partial_{r\phi}^{2}U+\partial^{2}_{ru}U+\partial_{r}U(-2U\partial_{\phi}\beta_{0}+3\partial_{u}\varphi-2\partial_{u}\beta_{0}+3U\partial_{\phi}\varphi))+e^{-2\beta_{0}-\varphi}(2\partial^{3}_{r\phi^2}V\nonumber\\
&-2r\partial^{2}_{r\phi}V(\partial_{\phi}\varphi-\partial_{\phi}\beta_{0})-6\partial^{2}_{\phi}V+6\partial_{\phi}V(\partial_{\phi}\varphi-\partial_{\phi}\beta_{0}))-4\mu r e^{-2\beta_{0}}(-r\partial^{2}_{r\phi}V-2r^2 U\partial^{2}_{\phi}\beta_{0}-\nonumber\\
&2r^{2}\partial^{2}_{u\phi}\beta_{0}+2\partial_{\phi}V+2\partial_{\phi}\beta_{0}(Ur^{2}\partial_{\phi}\varphi+r^{2}\partial_{u}\varphi+V))+re^{-4\beta_{0}+\varphi}(r(12r^2\partial_{\phi}U(\partial_{\phi}\varphi-\partial_{\phi}\beta_{0})+\nonumber\\
&9\partial_{\phi}V+r^2 U(\partial_{\phi}\beta_{0})^{2}+\partial_{\phi}\beta_{0}(-2Ur^{2}\partial_{\phi}\varphi+r^2\partial_{u}\beta_{0}-r^2\partial_{u}\varphi-4V)+r^2U(\partial_{\phi}\varphi)^{2}+\partial_{\phi}\varphi(4V-\nonumber\\
&8r^2\partial_{u}\beta_{0}+8r^2\partial_{u}\varphi))\partial_{r}U-12\partial_{u}V-r\partial^{2}_{r\phi}V(\partial_{r}U-4U)+12r\partial^{2}_{r\phi}U(-Ur^2\partial_{\phi}\beta_{0}+Ur^2\partial_{\phi}\varphi-\nonumber\\
&-2r^2\partial_{u}\beta_{0}+2r^2\partial_{u}\varphi+2r^2\partial_{\phi}U+V)+4r^{3}\partial^{2}_{ru}U(\partial_{\phi}\varphi-\partial_{\phi}\beta_{0})-r^{2}U\partial^{3}_{\phi r^{2}}V+r^{3}\partial_{r}U\partial^{2}_{u\phi}\varphi\nonumber\\
&+4r^3U\partial^{3}_{r\phi^2}U-r^3\partial^{2}_{u\phi}\beta_{0}\partial_{r}U+4r^3\partial^{2}_{\phi}U\partial_{r}U+2r^2V\partial^{2}_{\phi r^2}U-2r^2\partial^{2}_{r}V(-U\partial_{\phi}\beta_{0}+U\partial_{\phi}\varphi+\nonumber\\
&\partial_{u}\varphi-\partial_{u}\beta_{0}+\partial_{\phi}U)+r^2\partial^{2}_{r}U(3\partial_{\phi}V+V(\partial_{\phi}\varphi-\partial_{\phi}\beta_{0}))+12UV\partial_{\phi}\beta_{0})].
\end{align}

\section{Explicit terms of Lagrangian}\label{appexpl}
Here we provided the explicit terms of Lagrangian (\ref{totalact}).
\begin{align}
L_{bulk}=&-\dfrac{e^{2\beta_{0}+\varphi}R^2}{8\pi G l^2}+\dfrac{(2e^{2\beta_{0}}\partial_{\phi}\beta_{0}+(U_{0}\partial_{\phi}U_{0}+\partial_{u}U_{0})l^{2}e^{-2\beta_{0}+2\varphi})R}{32\pi G \mu l^2}+\dfrac{1}{16\pi G\mu} (U_{0}\partial^{2}_{\phi^2}\beta_{0}\nonumber\\
&+\partial_{\phi}U_{0}\partial_{\phi}\beta_{0}+\partial^{2}_{u\phi}\beta_{0})\ln(R)+\mathcal{O}\left(\dfrac{1}{R}\right),
\end{align}

\begin{align}
L_{GHY}=&-\dfrac{e^{2\beta_{0}+\varphi}R^2}{4\pi G l^2}-\dfrac{e^{\varphi}}{4\pi G}\left(\partial_{\phi}U_{0}+\partial_{u}\varphi+U_{0}\partial_{\phi}\varphi\right)R+\dfrac{1}{8\pi G}[e^{2\beta_{0}-\varphi}(4(\partial_{\phi}\beta_{0})^{2}+2\partial^{2}_{\phi}\beta_{0}-2\partial_{\phi}\beta_{0}\partial_{\phi}\varphi)\nonumber\\
&+l^{2}e^{2\beta_{0}-\varphi}(-U_{0}^2\partial^{2}_{\phi}\varphi-U_{0}\partial^{2}_{\phi}U_{0}-2\partial^{2}_{u\phi}\varphi U_{0}-\partial^{2}_{u\phi}U_{0}-\partial^{2}_{u}\varphi+2\partial_{\phi}\beta_{0}(U_{0}\partial_{u}\varphi+U_{0}\partial_{\phi}U_{0}+U_{0}^{2}\partial_{\phi}\varphi)\nonumber\\
&+\partial_{\phi}\varphi(2U_{0}\partial_{u}\beta_{0}-\partial_{u}U_{0}-U_{0}\partial_{\phi}U_{0})+2\partial_{u}\beta_{0}(\partial_{u}\varphi+\partial_{\phi}U_{0}))+V_{0}e^{\varphi}]+\mathcal{O}\left(\dfrac{1}{R}\right),
\end{align}

\begin{align}
L_{ct}=&-\dfrac{e^{2\beta_{0}+\varphi}R^2}{8\pi G l^2}-\dfrac{e^{\varphi}}{8\pi G}(\partial_{u}\varphi+U_{0}\partial_{\phi}\varphi +\partial_{\phi}U_{0})R+\dfrac{e^{\varphi -2\beta_{0}}}{16\pi G}(l^2 U_{0}^{2}(\partial_{\phi}\varphi)^2+2l^2 U_{0}\partial_{\phi}\varphi(\partial_{u}\varphi+\partial_{\phi}U_{0})\nonumber\\
&+l^{2}(\partial_{\phi}\varphi)^{2}+2l^2\partial_{u}\varphi \partial_{\phi}U_{0}+l^2(\partial_{\phi}U_{0})^{2}+V_{0}e^{2\beta_{0}})+\mathcal{O}\left(\dfrac{1}{R}\right),
\end{align}

\begin{align}
L_{0}=&\dfrac{e^{\varphi}(\partial_{u}\varphi+U_{0}\partial_{\phi}\varphi+\partial_{\phi}U_{0})R}{8\pi G}+\dfrac{e^{2\beta_{0}-\varphi}(\partial^{2}_{\phi^2}\beta_{0}-\partial_{\phi}\varphi \partial_{\phi}\beta_{0}+2(\partial_{\phi}\beta_{0})^2)}{4\pi G}+\mathcal{O}\left(\dfrac{1}{R}\right),
\end{align}

\begin{align}
L_{b}=&-\dfrac{l^2 e^{\varphi-2\beta_{0}}\left(\partial_{u}\varphi+U_{0}\partial_{\phi}\varphi+\partial_{\phi}U_{0}\right)^{2}}{16\pi G}+\mathcal{O}\left(\sqrt{\dfrac{1}{R}}\right),
\end{align}

\begin{align}
L_{\mathcal{R}}=&\dfrac{1}{8\pi G}[e^{2\beta_{0}-\varphi}(-4(\partial_{\phi}\beta_{0})^{2}+2\partial_{\phi}\beta_{0}\partial_{\phi}\varphi-2\partial^{2}_{\phi}\beta_{0})+\dfrac{e^{-2\beta_{0}+\varphi}l^2}{8\pi G}(U_{0}^{2}\partial^{2}_{\phi}\varphi+U_{0}\partial^{2}_{\phi}U_{0}+2U_{0}\partial^{2}_{u\phi}\varphi\nonumber\\
&+\partial^{2}_{u\phi}U_{0}+\partial^{2}_{u}\varphi+U_{0}^{2}(\partial_{\phi}\varphi)^{2}+\partial_{\phi}\varphi(-2U_{0}\partial_{u}\beta_{0}+\partial_{u}U_{0}+3U_{0}\partial_{\phi}U_{0}+2U_{0}\partial_{u}\varphi-2\partial_{\phi}\beta_{0}U_{0}^{2})\nonumber\\
&-2\partial_{\phi}\beta_{0}U_{0}(\partial_{u}\varphi+\partial_{\phi}U_{0})+(\partial_{\phi}U_{0})^{2}+2\partial_{\phi}U_{0}(\partial_{u}\varphi-\partial_{u}\beta_{0})+(\partial_{u}\varphi)^{2}-2\partial_{u}\beta_{0}\partial_{u}\varphi)]+\mathcal{O}\left(\dfrac{1}{R^{2}}\right),
\end{align}

\begin{equation}
L_{ccs}=2e^{2\beta_{0}}\partial_{\phi}\beta_{0}+l^2 e^{2\varphi-2\beta_{0}}\left(U_{0}\partial_{\phi}U_{0}+\partial_{u}U_{0}\right)
\end{equation}
\begin{equation}
L_{lccs}=\partial^{2}_{u\phi}\beta_{0}+\partial_{\phi}\beta_{0}\partial_{\phi}U_{0}+U_{0}\partial^{2}_{\phi^{2}}
\beta_{0}
\end{equation}

%\begin{align}
%L_{bcs}&=V_{0}U_{0}e^{2\varphi -2\beta_{0}}+U_{0}\partial^{2}_{\phi^{2}}\beta_{0}-U_{0}\partial^{2}_{\phi^{2}}\varphi +\partial^{2}_{u \phi}\beta_{0}-\partial^{2}_{ \phi}U_{0}-\partial^{2}_{u\phi}\varphi -\partial_{u}\varphi \partial_{\phi}\beta_{0}-\partial_{\phi}\varphi U_{0}\partial_{\phi}\beta_{0}\nonumber\\
%&+2U_{0}(\partial_{\phi}\beta_{0})^{2}+\partial_{\phi}U_{0}\partial_{\phi}\beta_{0}-2U_{0}\varphi \partial^{2}_{\phi^2}\beta_{0}+3U_{0}\partial^{2}_{\phi^2}\beta_{0}+2\beta_{0}U_{0}\partial^{2}_{\phi^2}\beta_{0}+
%2U_{0}\varphi \partial^{2}_{\phi^2}\varphi -2\beta_{0}U_{0}\partial^{2}_{\phi^2}\varphi \nonumber\\
%&-U_{0}\partial^{2}_{\phi^2}\varphi -U_{0}\partial_{\phi}\beta_{0}\partial_{\phi}\varphi -\partial_{\phi}U_{0}\partial_{\phi}\varphi-\partial_{u}\varphi\partial_{\phi}\varphi +2(\beta_{0}-\varphi)\partial^{2}_{u\phi}\beta_{0}+2(\beta_{0}-\varphi)(\partial^{2}_{u\phi}\beta_{0}-\partial^{2}_{\phi^2}U_{0}- \nonumber\\
%&\partial^{2}_{u\phi}\varphi -\partial_{\phi}\varphi \partial_{\phi}U_{0}+4\partial_{\phi}U_{0}\partial_{\phi}\beta_{0})
%+(\partial_{\phi}\beta_{0}-\partial_{\phi}\varphi)(\partial_{\phi}U_{0}-\partial_{u}\varphi)+2\partial_{\phi}\beta_{0}\partial_{u}\beta_{0}+6U_{0}\partial_{\phi}\beta_{0}^2
%+\nonumber\\
%&(2e^{2\beta_{0}}\partial_{\phi}\beta_{0}+(U_{0}\partial_{\phi}U_{0}
%+\partial_{u}U_{0})l^{2}e^{-2\beta_{0}+2\varphi})R+\mathcal{O}\left(\frac{1}{R}\right).
%\end{align}
The boundary Lagrangian (\ref{eqboundray}) is given as
\begin{align}
\bar{L}_{0}=&\dfrac{e^{-4\beta_{0}-\varphi}}{8\pi G}(l^2 e^{2\varphi}((U_{0}+U^{3}_{0})\partial_{\phi}\varphi+(1+3U^{2}_{0})\partial_{\phi}U_{0}-4(U_{0}+U^{3}_{0})\partial_{\phi}\beta_{0}-4(1+U^{2}_{0})\partial_{u}\beta_{0}\nonumber\\
&+\partial_{u}U^{2}_{0}+\partial_{u}\varphi+U^{2}_{0}\partial_{u}\varphi)+e^{4\beta_{0}}(U_{0}\partial_{\phi}\varphi-\partial_{\phi}U_{0})).
\end{align}

\section{Explicit terms of potentials}\label{appexpl1}
The renormalized presymplectic potential as
\begin{align}\label{thetaren}
\Theta^{r}_{ren}(g,\delta g)&=\dfrac{1}{16\pi G}[V_{0}e^{\varphi}\delta(\varphi-2\beta_{0})+2Ne^{\varphi}\delta U_{0}+2e^{2\beta_{0}-\varphi}(6\partial_{\phi}\beta_{0}\partial_{\phi}\delta\beta_{0}-\partial_{\phi}\varphi \partial_{\phi}\delta\beta_{0}+\partial^{2}_{\phi}\delta\beta_{0})]\nonumber\\
&+\dfrac{1}{64\pi \mu G}[-2V_{0}e^{2\varphi-2\beta_{0}}\delta U_{0}-2U_{0}\partial^{2}_{\phi}\delta\beta_{0}+2U_{0}\partial^{2}_{\phi}\delta\varphi-2\partial^{2}_{u \phi}\delta\beta_{0}+2\partial^{2}_{\phi}\delta U_{0}+2\partial^{2}_{u\phi}\delta\varphi-\nonumber\\
&\partial_{\phi}\delta\beta_{0}(-4U_{0}\partial_{\phi}\varphi-4\partial_{u}\varphi-2\partial_{\phi}U_{0})+\partial^{2}_{\phi}\beta_{0}(8U_{0}\delta\varphi-8U_{0}\delta\beta_{0}-6\delta U_{0})+\partial^{2}_{\phi}\varphi(-4U_{0}\delta\varphi\nonumber\\
&+4U_{0}\delta\beta_{0}+2\delta U_{0})+\partial^{2}_{\phi}\delta\varphi(4U_{0}\partial_{\phi}\beta_{0}-2\partial_{u}\varphi+2\partial_{\phi}U_{0})+8\partial^{2}_{u\phi}\beta_{0}\delta(\varphi-\beta_{0})-4\partial^{2}_{\phi}U_{0}\delta(\varphi-\beta_{0})\nonumber\\
&-4\partial^{2}_{u\phi}\varphi\delta(\varphi-\beta_{0})+\partial_{u}\delta\varphi(2\partial_{\phi}\varphi+4\partial_{\phi}\beta_{0})+\partial_{\phi}\varphi\partial_{\phi}\delta U_{0}+(\partial_{\phi}\varphi)^{2}(-2U_{0}\delta\varphi+2U_{0}\delta\beta_{0}+\delta U_{0})\nonumber\\
&+\partial_{\phi}\varphi(\delta\beta_{0}(-4U_{0}\partial_{\phi}\beta_{0}+2\partial_{u}\varphi+6\partial_{\phi}U_{0})+\delta\varphi(4U_{0}\partial_{\phi}\beta_{0}-2\partial_{u}\varphi-6\partial_{\phi}U_{0})+2\partial_{\phi}\beta_{0}\delta U_{0})-\nonumber\\
&4\partial_{\phi}\beta_{0}\delta\beta_{0}(\partial_{u}\varphi+2\partial_{\phi}U_{0})+4\partial_{\phi}\beta_{0}\delta\varphi(\partial_{u}\varphi+2\partial_{\phi}U_{0})-16\delta U_{0}(\partial_{\phi}\beta_{0})^{2}]+\mathcal{O}\left(r^{-1}\right),
\end{align}
the associated presymplectic current reads as
\begin{align}\label{eqwren}
w^{r}_{ren}(g,\delta_{1}g,\delta_{2}g)&=\dfrac{1}{16\pi G}[\delta_{2}(V_{0}e^{\varphi})\delta_{1}(\varphi-2\beta_{0})+2\delta_{2}(Ne^{\varphi})\delta_{1}U_{0}-2e^{2\beta_{0}-\varphi}\partial_{\phi}\delta_{2}\varphi \partial_{\phi}\delta_{1}\beta_{0}+\nonumber\\
&2e^{2\beta_{0}-\varphi}\delta_{2}(2\beta_{0}-\varphi)(6\partial_{\phi}\beta_{0}\partial_{\phi}\delta_{1}\beta_{0}-\partial_{\phi}\varphi\partial_{\phi}\delta_{1}\beta_{0}
+\partial^{2}_{\phi}\delta_{1}\beta_{0})]+\nonumber\\
&\dfrac{1}{64\pi G \mu}[-2\delta_{2}(V_{0}e^{2\varphi-2\beta_{0}})\delta_{1}U_{0}-
2\delta_{2}U_{0}\partial^{2}_{\phi}\delta_{1}\beta_{0}+2\delta_{2}U_{0}\partial^{2}_{\phi}\delta_{1}\varphi
+\nonumber\\
&(-4\delta_{2}(U_{0}\partial_{\phi}\varphi)-4\partial_{u}\delta_{2}\varphi-2\partial_{\phi}\delta_{2}U_{0})\partial_{\phi}\delta_{1}\beta_{0}+
8\delta_{2}(U_{0}\partial^{2}_{\phi}\beta_{0})\delta_{1}\varphi-8\delta_{2}(U_{0}\partial^{2}_{\phi}\beta_{0})\delta_{1}\beta_{0}\nonumber\\
&-6\partial^{2}_{\phi}\delta_{2}\beta_{0}\delta_{1} U_{0}
+4\delta_{2}(U_{0}\partial^{2}_{\phi}\varphi)\delta_{1}\varphi+4\delta_{2}U_{0}\delta_{1}\beta_{0}+\partial^{2}_{\phi}\delta_{1}
\varphi(4\delta_{2}(U_{0}\partial_{\phi}\beta_{0})-2\delta_{2}(\partial_{u}\varphi)\nonumber\\
&+2\delta_{2}(\partial_{\phi}U_{0}))+8\delta_{2}(\partial^{2}_{u\phi}\beta_{0})\delta_{1}(\varphi-\beta_{0})-4\delta_{2}(\partial^{2}_{\phi}U_{0})\delta_{1}(\varphi-\beta_{0})
-4\delta_{2}(\partial^{2}_{u\phi}\varphi)\delta_{1}(\varphi-\beta_{0})\nonumber\\
&+\partial_{u}\delta_{1}\varphi(2\delta_{2}(\partial_{\phi}\varphi)+4\delta_{2}(\partial_{\phi}\beta_{0}))+\partial_{\phi}\delta_{2}\varphi\partial_{\phi}\delta_{1} U_{0}-2\delta_{2}(U_{0}(\partial_{\phi}\varphi)^{2})\delta_{1}\varphi+2\delta_{2}(U_{0}(\partial_{\phi}\varphi)^{2})\delta_{1}\beta_{0}\nonumber\\
&+\delta_{2}((\partial_{\phi}\varphi)^{2})\delta_{1} U_{0}
+\partial_{\phi}\delta_{2}\varphi\delta_{1}\beta_{0}(-4U_{0}\partial_{\phi}\beta_{0}+2\partial_{u}\varphi+6\partial_{\phi}U_{0})+\partial_{\phi}\varphi\delta_{1}\beta_{0}(-4\delta_{2}U_{0}\partial_{\phi}\beta_{0})\nonumber\\
&+2\partial_{u}\delta_{2}\varphi+6\partial_{\phi}\delta_{2}U_{0})+\delta_{1}\varphi(4\delta_{2}(U_{0}\partial_{\phi}\beta_{0})-2\partial_{u}\delta_{2}\varphi-6\partial_{\phi}\delta_{2}U_{0})+2\partial_{\phi}\delta_{2}\beta_{0}\delta_{1} U_{0})-\nonumber\\
&4\partial_{\phi}\beta_{0}\delta_{1}\beta_{0}(\partial_{u}\delta_{2}\varphi+2\partial_{\phi}\delta_{2}U_{0})+4\partial_{\phi}\delta_{2}\beta_{0}\delta_{1}\varphi(\partial_{u}\varphi+2\partial_{\phi}U_{0})+4\partial_{\phi}\beta_{0}\delta_{1}\varphi(\partial_{u}\delta_{2}\varphi+2\partial_{\phi}\delta_{2}U_{0})\nonumber\\
&-16\delta_{1} U_{0}\delta_{2}((\partial_{\phi}\beta_{0})^{2})]-(1\leftrightarrow 2)+\mathcal{O}(r^{-1}).
\end{align}

The explicit expression for the infinitesimal charges (\ref{eqintg}) reads as

\begin{align}\label{eqchar}
\ndelta Q_{\xi}[g]&=\int_{0}^{2\pi}d\phi k^{u r}_{ren,\xi}[g;\delta g]=\dfrac{1}{8\pi G}\int_{0}^{2\pi}d\phi [Y\delta(N e^{\varphi})+\partial_{\phi}(e^{2\beta_{0}-\varphi}\partial_{\phi}f)\delta(\beta_{0}-\varphi)+f(\dfrac{1}{2}\delta V_{0}e^{\varphi}\nonumber\\
&-V_{0}e^{\varphi}\delta(\beta_{0}-\varphi)-U_{0}\delta(Ne^{\varphi})+2e^{2\beta_{0}-\varphi}(6\partial_{\phi}\beta_{0}\partial_{\phi}\delta\beta_{0}-\partial_{\phi}\varphi \partial_{\phi}\delta\beta_{0}+\partial^{2}_{\phi}\delta\beta_{0}))]+\dfrac{1}{4\pi G \mu}[\nonumber\\
&(Y-fU_{0})\partial^{2}_{\phi}\delta\beta_{0}+((7Y-6fU_{0})\partial_{\phi}\beta_{0}+(fU_{0}-\dfrac{3}{2}Y)\partial_{\phi}\varphi+\dfrac{1}{2}(\partial_{u}f-\partial_{\phi}Y-w+2U_{0}\partial_{\phi}f\nonumber\\
&+2\partial_{u}\beta_{0} f-f\partial_{u}\varphi))\partial^{2}_{\phi}\delta\beta_{0}+(Y-2fU_{0})\partial^{2}_{\phi}\beta_{0}-\dfrac{1}{2}(Y-2fU_{0})\partial^{2}_{\phi}\varphi-f\partial^{2}_{\phi u}\beta_{0}+f\partial^{2}_{\phi}U_{0}\nonumber\\
&+\dfrac{f}{2}\partial^{2}_{\phi u}\varphi-\dfrac{1}{2}\partial^{2}_{\phi}Y-\dfrac{1}{2}\partial^{2}_{u \phi}f+(2Y(\partial_{\phi}\beta_{0})^{2}-\partial_{\phi}\beta_{0}(2\partial_{\phi}Y-2f\partial_{\phi}U_{0}+Y\partial_{\phi}\varphi-\partial_{u}f+f\partial_{u}\varphi\nonumber\\
&-2f\partial_{u}\beta_{0}-4U_{0}\partial_{\phi}f+w)-\partial_{\phi}\varphi(U_{0}\partial_{\phi}f-\dfrac{1}{2}\partial_{\phi}Y+f\partial_{\phi}U_{0})-\partial_{\phi}f\partial_{\phi}U_{0}-\dfrac{1}{2}\partial_{\phi}w-\dfrac{1}{2}\partial_{\phi}f\partial_{u}\varphi\nonumber\\
&+\partial_{\phi}f\partial_{u}\beta_{0})\delta\beta_{0}+\dfrac{\delta\varphi}{2}(-2Y(\partial_{\phi}\beta_{0})^{2}+\partial_{\phi}\beta_{0}(-2\partial_{u}\beta_{0}f-4f\partial_{\phi}U_{0}+Y\partial_{\phi}\varphi+3\partial_{\phi}Y+w+f\partial_{u}\varphi\nonumber\\
&-6U_{0}\partial_{\phi}f-\partial_{u}f)+\partial_{\phi}\varphi(2U_{0}\partial_{\phi}f-\partial_{\phi}Y+2f\partial_{\phi}U_{0})+
2\partial_{\phi}f\partial_{\phi}U_{0}+\partial_{\phi}w+\partial_{\phi}f\partial_{u}\varphi-2\partial_{\phi}f\partial_{u}\beta_{0})\nonumber\\
&+(-V_{0}\delta\beta_{0}+\dfrac{1}{2}\delta V_{0}+V_{0}\delta\varphi)(Y-fU_{0})e^{2\varphi-2\beta_{0}}].
\end{align}


\begin{thebibliography}{99}


\bibitem{Bondi:1962px}
H.~Bondi, M.~G.~J.~van der Burg and A.~W.~K.~Metzner,
%``Gravitational waves in general relativity. 7. Waves from axisymmetric isolated systems,''
Proc. Roy. Soc. Lond. A \textbf{269}, 21-52 (1962)
doi:10.1098/rspa.1962.0161
%1349 citations counted in INSPIRE as of 04 Mar 2021


\bibitem{Barnich:2007bf}
G.~Barnich and G.~Compere,
%``Surface charge algebra in gauge theories and thermodynamic integrability,''
J. Math. Phys. \textbf{49}, 042901 (2008)
doi:10.1063/1.2889721
[arXiv:0708.2378 [gr-qc]].
%214 citations counted in INSPIRE as of 29 Jul 2021

\bibitem{Wald:1999wa}
R.~M.~Wald and A.~Zoupas,
%``A General definition of 'conserved quantities' in general relativity and other theories of gravity,''
Phys. Rev. D \textbf{61}, 084027 (2000)
doi:10.1103/PhysRevD.61.084027
[arXiv:gr-qc/9911095 [gr-qc]].
%390 citations counted in INSPIRE as of 29 Jul 2021

\bibitem{Ruzziconi:2020wrb}
R.~Ruzziconi and C.~Zwikel,
%``Conservation and Integrability in Lower-Dimensional Gravity,''
JHEP \textbf{04}, 034 (2021)
doi:10.1007/JHEP04(2021)034
[arXiv:2012.03961 [hep-th]].
%9 citations counted in INSPIRE as of 28 Jul 2021

\bibitem{Barnich:2010eb}
G.~Barnich and C.~Troessaert,
%``Aspects of the BMS/CFT correspondence,''
JHEP \textbf{05} (2010), 062
doi:10.1007/JHEP05(2010)062
[arXiv:1001.1541 [hep-th]].
%351 citations counted in INSPIRE as of 13 Feb 2021


\bibitem{Barnich:2013axa}
G.~Barnich and C.~Troessaert,
%``Comments on holographic current algebras and asymptotically flat four dimensional spacetimes at null infinity,''
JHEP \textbf{11}, 003 (2013)
doi:10.1007/JHEP11(2013)003
[arXiv:1309.0794 [hep-th]].
%73 citations counted in INSPIRE as of 04 Mar 2021

\bibitem{Adami:2020ugu}
H.~Adami, M.~M.~Sheikh-Jabbari, V.~Taghiloo, H.~Yavartanoo and C.~Zwikel,
%``Symmetries at null boundaries: two and three dimensional gravity cases,''
JHEP \textbf{10}, 107 (2020)
doi:10.1007/JHEP10(2020)107
[arXiv:2007.12759 [hep-th]].
%11 citations counted in INSPIRE as of 04 Mar 2021

\bibitem{ban}M. Banados, C. Teitelboim and J. Zanelli, Phys.
Rev. Lett. 69, 1849 (1992).

\bibitem{BrownHenneaux}
J.~D.~Brown and M.~Henneaux, Commun.\ Math.\ Phys.\  {\bf 104}, 207 (1986).

\bibitem{jak}S. Deser, R. Jackiw and S. Templeton, Phys. Rev. Lett.
48, 975 (1982); Annals Phys. 140, 372 (1982).

\bibitem{Skenderis:2009nt}
K.~Skenderis, M.~Taylor and B.~C.~van Rees,
%``Topologically Massive Gravity and the AdS/CFT Correspondence,''
JHEP \textbf{09} (2009), 045
doi:10.1088/1126-6708/2009/09/045
[arXiv:0906.4926 [hep-th]].
%129 citations counted in INSPIRE as of 25 Apr 2021

\bibitem{Adami:2021sko}
H.~Adami, M.~M.~Sheikh-Jabbari, V.~Taghiloo, H.~Yavartanoo and C.~Zwikel,
%``Chiral Massive News: Null Boundary Symmetries in Topologically Massive Gravity,''
[arXiv:2104.03992 [hep-th]].
%0 citations counted in INSPIRE as of 22 May 2021


\bibitem{Flanagan:2015pxa}
\'E.~\'E.~Flanagan and D.~A.~Nichols,
%``Conserved charges of the extended Bondi-Metzner-Sachs algebra,''
Phys. Rev. D \textbf{95}, no.4, 044002 (2017)
doi:10.1103/PhysRevD.95.044002
[arXiv:1510.03386 [hep-th]].
%94 citations counted in INSPIRE as of 04 Mar 2021

\bibitem{Penrose:1965am}
R.~Penrose,
%``Zero rest mass fields including gravitation: Asymptotic behavior,''
Proc. Roy. Soc. Lond. A \textbf{284}, 159 (1965)
doi:10.1098/rspa.1965.0058
%479 citations counted in INSPIRE as of 04 Mar 2021


\bibitem{Compere:2020lrt}
G.~Comp\`ere, A.~Fiorucci and R.~Ruzziconi,
%``The $\Lambda$-BMS$_4$ charge algebra,''
JHEP \textbf{10}, 205 (2020)
doi:10.1007/JHEP10(2020)205
[arXiv:2004.10769 [hep-th]].
%19 citations counted in INSPIRE as of 04 Mar 2021


\bibitem{rig}
M. Riegler and C. Zwikel, PoS Modave2017
(2018) 004.

\bibitem{Ciambelli:2020eba}
L.~Ciambelli, C.~Marteau, P.~M.~Petropoulos and R.~Ruzziconi,
%``Gauges in Three-Dimensional Gravity and Holographic Fluids,''
JHEP \textbf{11} (2020), 092
doi:10.1007/JHEP11(2020)092
[arXiv:2006.10082 [hep-th]].
%4 citations counted in INSPIRE as of 13 Feb 2021


\bibitem{Iyer:1994ys}
V.~Iyer and R.~M.~Wald,
%``Some properties of Noether charge and a proposal for dynamical black hole entropy,''
Phys. Rev. D \textbf{50} (1994), 846-864
doi:10.1103/PhysRevD.50.846
[arXiv:gr-qc/9403028 [gr-qc]].
%1498 citations counted in INSPIRE as of 24 Apr 2021

\bibitem{leewald}
J. Lee and R. M. Wald, “Local symmetries and constraints,” J. Math. Phys. 31 (1990) 725–743.

\bibitem{Barnich:2011mi}
G.~Barnich and C.~Troessaert,
%``BMS charge algebra,''
JHEP \textbf{12} (2011), 105
doi:10.1007/JHEP12(2011)105
[arXiv:1106.0213 [hep-th]].
%253 citations counted in INSPIRE as of 25 Apr 2021


\bibitem{Dengiz:2020fpe}
S.~Dengiz, E.~Kilicarslan and M.~R.~Setare,
%``Lee-Wald Charge and Asymptotic Behaviors of the Weyl-invariant Topologically Massive Gravity,''
Class. Quant. Grav. \textbf{37} (2020) no.21, 215016
doi:10.1088/1361-6382/abbc46
[arXiv:2002.00345 [hep-th]].
%0 citations counted in INSPIRE as of 25 Apr 2021


\bibitem{deHaro:2000vlm}
S.~de Haro, S.~N.~Solodukhin and K.~Skenderis,
%``Holographic reconstruction of space-time and renormalization in the AdS / CFT correspondence,''
Commun. Math. Phys. \textbf{217} (2001), 595-622
doi:10.1007/s002200100381
[arXiv:hep-th/0002230 [hep-th]].
%1425 citations counted in INSPIRE as of 25 Apr 2021

\bibitem{Ciambelli:2020ftk}
L.~Ciambelli, C.~Marteau, P.~M.~Petropoulos and R.~Ruzziconi,
%``Fefferman-Graham and Bondi Gauges in the Fluid/Gravity Correspondence,''
PoS \textbf{CORFU2019}, 154 (2020)
doi:10.22323/1.376.0154
[arXiv:2006.10083 [hep-th]].
%6 citations counted in INSPIRE as of 23 May 2021

%\cite{Detournay:2014fva}
\bibitem{Detournay:2014fva}
S.~Detournay, D.~Grumiller, F.~Sch\"oller and J.~Sim\'on,
%``Variational principle and one-point functions in three-dimensional flat space Einstein gravity,''
Phys. Rev. D \textbf{89}, no.8, 084061 (2014)
doi:10.1103/PhysRevD.89.084061
[arXiv:1402.3687 [hep-th]].
%45 citations counted in INSPIRE as of 10 Sep 2021


%\cite{Compere:2020lrt}
\bibitem{Compere:2020lrt}
G.~Comp\`ere, A.~Fiorucci and R.~Ruzziconi,
%``The $\Lambda$-BMS$_4$ charge algebra,''
JHEP \textbf{10}, 205 (2020)
doi:10.1007/JHEP10(2020)205
[arXiv:2004.10769 [hep-th]].
%41 citations counted in INSPIRE as of 10 Sep 2021

\end{thebibliography}
\end{document}